# Photonics study of probiotic treatment on brain cells exposed to chronic alcoholism using molecular specific nuclear light localization properties via confocal imaging

Prakash Adhikari,[1] Pradeep K. Shukla,[2] Mehedi Hasan,[1] Fatemah Alharthi,[1] Binod Regmi,[1] and Radhakrishna Rao,[2] Prabhakar Pradhan [1,*]

[1] Department of Physics and Astronomy, Mississippi State University, Mississippi State, MS, USA, 39762
[2] Department of Physiology, University of Tennessee Health Science Center, Memphis, TN, USA, 38103
pp838@msstate.edu and pshukla2@uthsc.edu and

**Abstract:** Molecular specific photonics localization technique, the inverse participation ratio (IPR), is a powerful technique to probe the nanoscale structural alterations due to abnormalities or chronic alcoholism in brain cells using the confocal image. Chronic alcoholism is correlated with medical, behavioral, and psychological problems including brain cell damage. However, probiotics such as *Lactobacillus Plantarum* has shown the promising result in soothing the human brain. This report, using the Confocal-IPR technique, nano to submicron scale structural abnormalities of the glial cells and the nuclei of alcoholic mice brain in the presence of probiotics. The increase in the structural disorder of alcoholic brain cells while the decrease or normalcy in the structural disorder of brain cells of mice fed with probiotics and alcohol simultaneously indicates that alcohol stimulates probiotics and enhances brain function.

## 1. Introduction

Photonics/light is an important probe for the characterization of the structural properties of cells. The structural properties of cells and tissues change from the normal with the progress of disease or abnormalities, such as cancer, stress, drugs, etc. Cells and tissues are weakly disordered dielectric media, therefore their structure, as well as any change in the structural properties, can be characterized by using light. The light probing is generally done by measuring the scattering signals to quantify the optical parameters of the samples. Different types of structural changes can happen in a cell with the progression of diseases such as carcinogenesis or due to the effects of drugs[1,2]. These changes in a cell may range from bulk structural change to nanoscale molecular specific structures alterations. The nanoscale structural changes in cells/tissues due to diseases or abnormalities are associated with the mass density fluctuation, in turn, the refractive index fluctuations that are explored using recently introduced mesoscopic-physics based sensitive imaging technique called the partial wave spectroscopy (PWS) [2–6]. Although the overall change in a cell/nucleus has been studied for a long time to probe structural changes in a cell, however, these changes appear at the later stages of abnormalities or diseases which may be too late for treatment. The significant changes occur at the early stages of the abnormalities or diseases, where the molecular specific nanoscale structural changes occur due to the rearrangements of the macromolecules present in a cell, which is the major interest of study. Therefore, we hypothesize that probing the molecular specific structural changes can explain the physical state of a system. The concept of characterizing the molecular specific structural changes has been introduced recently to quantify the nanoscale changes in cells in abnormalities [1,3,7]. The initial case studies ranging from cancer to abnormalities in brain cells due to chronic alcoholism show promising results to characterize the physical state of the cell [7–9].

Chronic alcoholism affects the brain, in cells and tissue levels. It has been shown by transmission emission microscopy (TEM) imaging that structural properties of the cells changes in chronic alcoholism at the nanoscales, and changes are prominent at around the length scales of ~100nm, which is related to the building blocks of the cells. Glial cells such as astrocytes, and microglia, and chromatin of the brain cells are specifically found to be highly affected due to chronic alcoholism which totals the vital portion of the central nervous system (CNS). Significance of literature showed that alcohol consumption damages the structure of the brain, from cellular to the molecular level. Chronic alcoholism can produce sustainable damage in the brain. This structural damage can bring neuropathological disorders, such as memory loss, dysfunction of the brain resulting in cognitive and behavioral deficits [10–12].

In particular, astrocytes are star-shaped and most abundant cells i.e. glial cells in the central nervous system (CNS) and largely play as a regulator in the CNS immunity. In the same way, chromatin in all brain cell nuclei is a component that plays important role in genetic inheritance. In humans, a single astrocyte can interact with two million neurons at a time [13]. These astrocytes perform various tasks such as axon guidance, synoptic support, controlling blood flow and blood brain barrier as well. Along with their regulating behavior, they play a vital role in neuroinflammation in both beneficial and detrimental ways depending on the stimuli they receive from their inflamed environment. Alcohol is one of the well-known detrimental stimuli for neuroinflammation. Pathogenesis of many CNS disorders and several neurodegenerative diseases are crucially caused by brain inflammation [14]. This inflammation activates the glial cell, mainly the microglia and astrocytes that liberate the free radicals, cytokines, and inflammatory mediators that can damage the normal brain function [15]. A recent study showed that alcohol treatment in the microglial cells and chromatin altered their morphology [16].

In some cases, these structural changes may be reversible by different types of treatments. It has been reported that probiotic treatment has a good effect on brain health. In particular, it can reduce brain structural damage to a certain extent [17,18]. Initial results also show that probiotic treatment such as *Lactobacillus Plantarum* can be more effective in the presence of alcohol. *Lactobacillus Plantarum* is a Gram-positive lactic acid bacterium present in fermented food and in the gastro intestinal tract which has ample application in the medical field. The recently developed mesoscopic physics-based light localization technique, inverse participation ratio (IPR), can quantify the structural change in cells/tissues as the degree of structural disorder due to diseases or any other abnormalities [1,3,8,9]. In the IPR technique, an optical lattice is constructed using the pixel intensities of the confocal image and solved it for eigenvalues and eigenfunction under the closed boundary conditions. Finally, the eigenfunction is used to calculate the light localization properties of samples by the average of the IPR, *<IPR>* and the standard deviation of the IPR, $\sigma(<IPR>)$. The earlier IPR analysis results show that the degree of disorder strength i.e. $L_d$ is proportional to *<IPR>* or $\sigma(IPR)$ [7,8]. This technique has been used to quantify the extent of aggressive cancer in biological cells, especially to detect the progress of carcinogenesis and to study the anti-cancerous drug effect [8,19]. We want to extend this method now to probe the molecular specific spatial structural changes in brain cells and organelles due to chronic alcoholism.

In this work, using the IPR technique on a confocal image, we will first study the effect of alcohol in glial cells astrocytes, and microglia, and nuclei of mice brain cells by probing molecular specific structural changes at the nanoscale level. Then the effect of probiotic treatment in astrocytes, microglia, and nuclei in the brain cells of control fed, and lastly, the effect of probiotic in chronic alcoholism is studied.

## 2. Method

*2.1 Molecular specific structural disorder analysis of confocal images by applying the inverse participation ratio (IPR) analysis technique:*

Confocal imaging is a technique used to capture the images with high optical contrast and resolution, ranging at submicron length scales. The main principle of the confocal microscopy imaging is that a spatial pinhole is used to block out-of-focus light to acquire a controlled depth of field and reduced background lights in images. Here, the samples are first treated with fluorescence that can emit a broadband light intensity at a specific wavelength. The amount of fluorophore dye that binds a molecular mass is proportional to the molecular mass at any point or a small voxel volume. Therefore, the fluorescence light collected at any point is approximately proportional to the fluorophore dye/elements, specific molecular mass present at that point or voxel. As dyes are independent of each other, and treating a cell with different molecular binding dyes at the same time and then probing appropriate wavelengths can provide us the different molecular specific spatial structural mass density fluctuations in a cell.

The nanoscale mass density fluctuations can be quantified by calculating the degree of structural disorders in confocal images using the inverse participation ratio (IPR) technique. In the IPR technique, an optical lattice is formed using the pixel intensities of the confocal images. The optical lattice is a representation of the 'mass density fluctuations' that are scanned voxel-wise. Then Anderson tight binding model is used to obtain the Hamiltonian of the closed system and the eigenfunctions of optical lattices of light waves are used to analyze the localization properties of the sample. The light localization strength, therefore, indicates the level of structural disorder in the abnormal cells. An increase in the mass density fluctuations in the cells at the nanoscale level is represented by a higher value of the IPR. The *<IPR>* or $\sigma(<IPR>)$ ultimately quantify the degree of structural disorder in the medium.

As mentioned in [7,8], the pixel intensity $I(r)$ of the confocal image at position $r$ and $\rho$ is the density of molecule at voxel amount $dV(dV=dxdydz)$ of the image can be defined as:

$$I(r) \propto dV(\rho) \qquad (1)$$

It has been shown in [8,20] that the local refractive index of the cell slice is directly proportional to the local mass density of the cell at that point *(x,y)* and can be written as:

$$n(x, y) = n_0 + dn(x, y) \tag{2}$$

where $n_o$ is the average refractive index of the confocal images, $dn(x,y)$ is the refractive index fluctuation at position $(x,y)$ of the voxel $dV$.

Now, If $I_0$ is the average pixel intensity of the confocal images over the surface and $dI(x,y)$ is the intensity fluctuation of the pixel at position $(x, y)$ of the sample. Then, $I(x,y)$ represents the confocal image intensity at any voxel point $(x,y)$ of the sample cell given as:

$$I(x, y) = I_0 + dI(x, y) \tag{3}$$

It is to be noted that the value of intensity fluctuation, $dI(x,y)$ is always less than the average intensity $I_0$ ($dI << I_0$). In the same way, the refractive index fluctuation, $dn(x,y)$ is also less than the average refractive index $n_0$ ($dn << n_0$).

The refractive index $n(x,y)$ of thin scattering substances such as biological cells has a linear relation to the mass density [3,21]. The pixel intensity values at position $(x, y)$ can be correlated with the refractive index fluctuation due to the mass density variation of the fluorescence molecules. Therefore, the confocal image's intensity $I(x, y)$ is linearly proportional to the mass density $M(x, y)$, and refractive index $n(x, y)$ of the voxel. Also, a representative refractive index matrix constructed using the pixel intensity value [22] can be correlated with optical potential $\varepsilon_i(x,y)$ as:

$$\varepsilon_i(x, y) = \frac{dn(x, y)}{n_0} \propto \frac{dI(x, y)}{I_0} \tag{4}$$

The pixel intensity values in the confocal image give the onsite optical potential, $\varepsilon_i(x, y)$ which is a representation of the spatial refractive index fluctuations of the fluorescent molecules inside the sample [21,23].

Anderson's tight binding model (TBM) is a widely recognized model studied in condensed matter physics and can be used to explain the disorder properties of any optical geometrical systems [8]. The Hamiltonian approach of the Anderson Tight Binding Model (TBM) produces the eigenfunctions to analyze the spatial structural properties of an optical lattice that has been produced from the confocal image. The optical potential at every point can be obtained from equation (4). If we consider one optical state per lattice site and the inner lattice site hopping restricted to the nearest neighbors then the Hamiltonian of Anderson tight-binding model [22,24,25] generated as:

$$H = \sum_i \varepsilon_i |i\rangle\langle j| + t \sum_{<ij>} (|i\rangle\langle j| + |j\rangle\langle i|) \tag{5}$$

Here, $\varepsilon_i(x,y)$ is the optical potential energy of the $i^{th}$ lattice site, $|i>$ and $|j>$ are the eigenvectors of the $i^{th}$ and the $j^{th}$ lattice sites, and $t$ is the inter-lattice site hopping strength.

By the diagonalization method, we can generate the eigenfunctions ($E_i$'s) from the Hamiltonian, $H$ of the system. Using these eigenfunctions ($E_i$'s) we calculate the average IPR value, $<IPR>$ of the entire sample images as defined in [1,3,26,27]:

$$\langle IPR \rangle = \frac{1}{N} \sum_{i=1}^{N} \int_0^L \int_0^L E_i^4(x, y) dx dy \tag{6}$$

Where $E_i$ is the $i^{th}$ eigenfunction of the Hamiltonian $H$ of the optical lattice size $L \times L$, and $N$ is the total number of potential points on the refractive index matrix.

For the heterogeneous light transparent medium such as biological cells, two parameters namely the refractive index fluctuations $dn$ and its spatial fluctuation correlation length $l_c$ are used to specify the disorder of the system. This collective measurement provides the refractive index fluctuations inside the system termed as the disorder strength, $L_d$ [7,8,23]. It is shown in [7,8] that the average IPR value $<IPR>$ is directly proportional to the structural disorder strength i.e. $L_d =< dn > \times l_c$, and can be expressed as:

$$\langle IPR \rangle \propto L_d =< dn > \times l_c \tag{7}$$

$$\sigma(\langle IPR \rangle) \propto L_d =< dn > \times l_c \tag{8}$$

Lastly, the average ($<IPR>$) and standard deviation ($\sigma(<IPR>)$) of the IPR values or $L_d$ indicated as potential biomarkers of the confocal images of biological cells to understand the morphological condition under any abnormalities.

*2.2 Brain Cell Samples Preparation using a mouse model:*

Here, we study the effect of probiotics on chronic alcoholic's brain cells and nuclei using a mouse model. In particular, we have studied the glial cells: astrocytes, and microglia which include the major portion of CNS, molecular structural properties of astrocytes cells using glial fibrillary acidic protein (GFAP) antibody and microglia cells using Anti-TMEM119 transmembrane protein antibody; DNA molecular structural properties of chromatin structure in the nuclei of brain cells using DAPI.

To study the effect of alcohol and probiotics in the mice brain cells, 8-10 weeks mice bought from Jackson Laboratory were divided into 3-5 mice into 4 different groups. They are divided into four groups, equal males and females: (1) Control (PF), (2) Ethanol Fed (EF), (3) Probiotic Fed (PF+LP), (4) Probiotic and Alcohol Fed (EF+LP). To be more specific, we observed two glial cells and nuclei of the brain cells, namely: (I) Astrocytes, (II) Microglia, (III) Chromatin.

The rationale for choosing the mentioned glial cells and chromatin. (I) Astrocytes: These cells represent the major portion of the CNS and can interact millions of neurons at a time. They get exposed to the external simulator such as alcohol and probiotic. This can be easily tagged by the glial fibrillary acidic protein (GFAP) antibody marker. (II) Microglia: These cells, similar to the astrocytes are the major part of CNS system and considered as the macrophages of the CNS which clean up the cellular debris and participate in neuroinflammation. They show their over-expression due to the presence of stimulators like alcohol and probiotic, and the cell can be tagged by the dye called Anti-TMEM119 transmembrane protein antibody. (III) Chromatin: In all brain cell nuclei, chromatin is a major part. Chromatin is the main DNA molecular component of the cell. It has been shown that chromatin spatial structures changes with the progress of disease and any abnormalities. A previous study using TEM probing has shown that nanoscale structural disorder increases with the progress of alcoholism. The chromatin can be easily stained by the standard DAPI dye. The nanoscale changes in these types of cells are shown well for the detection of cancer stages. We hypothesize that this will also work well for the brain disorder case.

*Bio-Marker:* The main marker is to probe the confocal images using the IPR technique to quantify the molecular specific nanoscale molecular spatial mass density fluctuations, as mentioned in section 2.1.

As described above mice were randomly divided into 4 categories and fed ethanol in a standard protocol in Leiber DiCarli liquid diet (0-6% v/v stepwise increase) with or without *Lactobacillus Plantarum* ($10^6$ cfu/ml) for 4 weeks. (I) The first group was saline fed or vehicle (II) the second group was fed with ethanol (III) the third group was fed with the probiotic only (IV) fourth group was fed with probiotics in the presence of ethanol.

*Brain tissues:* When the mice were treated for 4 weeks, they were sacrificed and brains were removed. The brains were then cryo freeze and sectioned to 10μm slices by a microtome.

*Staining:* The different sections of the brain were treated with three types of dyes: GFAP antibody, Anti-TMEM119 antibody and DAPI.

*Confocal imaging:* Brain cryo sections (10μm thick) were fixed in acetone: methanol (1:1) at −20°C for 2 min and rehydrated in PBS (137 mM sodium chloride, 2.7 mM potassium chloride, 10 mM disodium hydrogen phosphate, and 1.8 mM potassium dihydrogen phosphate). Sections were permeabilized with 0.2% Triton X-100 in PBS for 10 min and blocked in 4% nonfat milk in Triton-Tris buffer (150 mM sodium chloride containing 10% Tween 20mM and 20mM Tris, pH 7.4). It was then incubated for 1 hour with the primary antibodies (mouse monoclonal anti- Alexa Fluor 488–conjugated GFAP and rabbit polyclonal anti–TMEM119), followed by incubation for 1 hour with secondary antibodies Cy3-conjugated anti-rabbit IgG antibodies) and 10 min incubation with Hoechst 33342. The fluorescence was examined using a Zeiss 710 confocal microscope (Carl Zeiss GmbH, Jena, Germany), and images from $x$–$y$ sections (1μm) were collected by LSM 5 Pascal software (Carl Zeiss Microscopy). Images were stacked by ImageJ software (Image Processing and Analysis in Java; National Institutes of Health, Bethesda, MD, USA) and processed by Adobe Photoshop (Adobe Systems, San Jose, CA, USA). All images for tissue samples from different group were collected and processed under identical conditions.

*IPR analyses:* IPR analyses were performed to quantify the nanoscale structural change in the cells and nuclei, as discussed above.

### 3. Results

Confocal images of the mice brain cells stained with different protein/dye were obtained to study the effect of alcohol and probiotics in astrocytes, microglia and nuclei of the mice brain cells. IPR analyses were performed on the individual confocal images at $(l \times l)nm^2$ as described. The degree of $<IPR(l)>$ values at different length scales for all molecular specific components of the mice brain cells were calculated. At each length scale $l$, at least 5 confocal micrographs of each category: astrocytes, microglia, and nuclei, of mice brain cells were analyzed separately and the ensemble averaging *IPR* calculated for their STD values. Finally, the calculated *std(<IPR>)* of astrocytes, microglia, and nuclei of the mice's brain cells of control, ethanol, and probiotics and ethanol fed were compared.

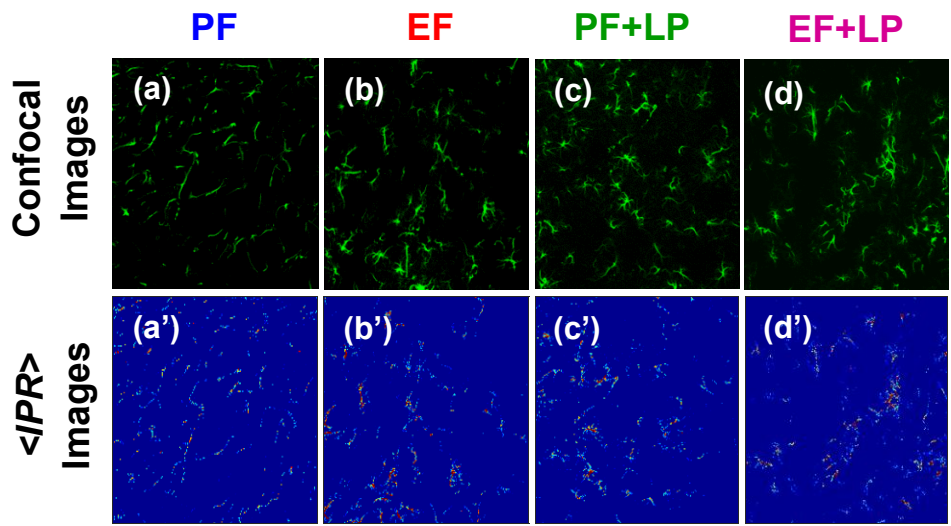

**Fig. 1: Confocal and IPR images of astrocytes (PF, EF, PF+LP, EF+LP): (a)-(d)** are the representative confocal images of control (PF), ethanol fed(EF), probiotics fed(PF+LP), and ethanol and probiotics fed(EF+LP) astrocytes of mice brain cells while **(a')-(d')** are their corresponding IPR images respectively. The IPR images are distinct from the confocal images and shows that structural disorder in astrocytes increases in the presence of ethanol as well as probiotics, but when treated together they decreases the alcohol damage.

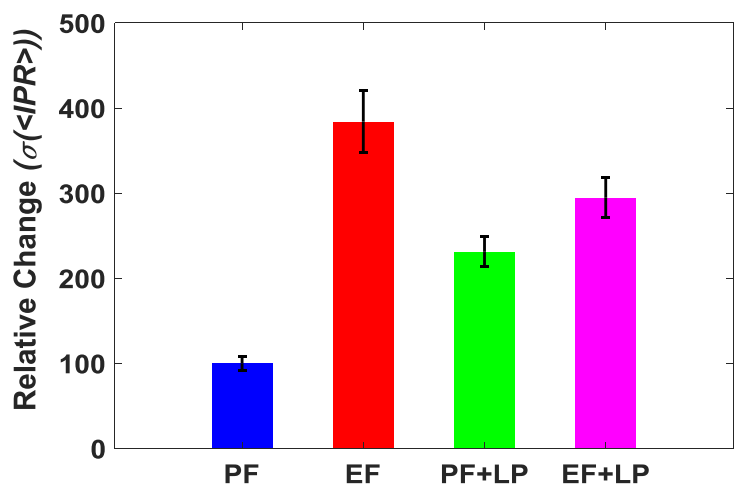

**Fig. 2:** Bar graph represents the relative study of molecular specific light localization property or $\sigma(<IPR>)$ of astrocytes, brain cells of control (PF) mice with respect to ethanol fed (EF), probiotic fed (PF+LP), and ethanol and probiotics fed (EF+LP) mice. The IPR analysis of astrocytes show that the *std* of disorder strength or $\sigma(<IPR>)$ of EF mice increase by 280% relative to PF mice. The $\sigma(<IPR>)$ of astrocytes cells of mice fed with only probiotics is relatively higher compared to the control mice. This increase in aggressiveness indicates that there is some effect of probiotics in the astrocytes. However, the $\sigma(<IPR>)$ of astrocytes cells of mice fed with EF+LP at the same time decreases by 100% relative to ethanol fed mice. This implies probiotics in the presence of ethanol get more stimulated and are beneficial for brain cells, reducing the $\sigma(<IPR>)$ of astrocytes of the brain cells.

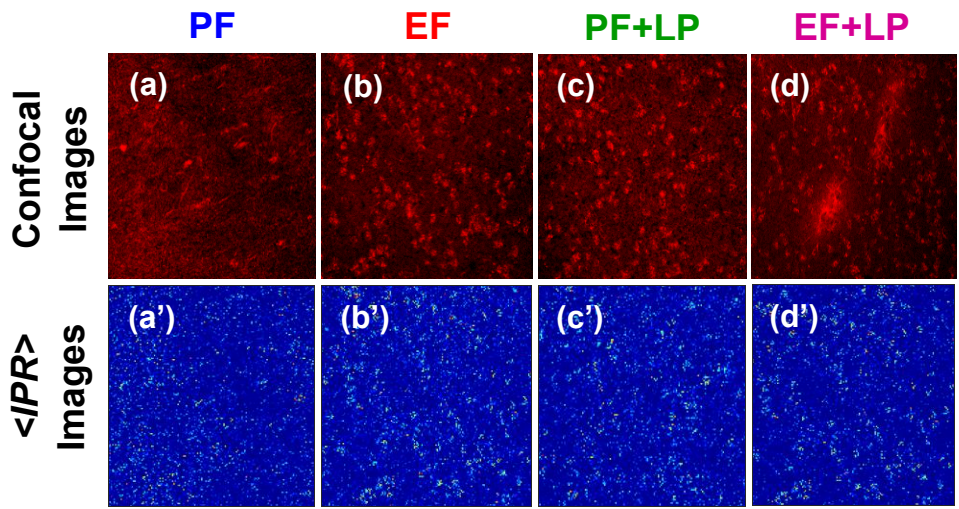

**Fig: 3. Confocal and IPR images of microglia (PF, EF, PF+LP, EF+LP): (a)-(d)** are the representative confocal images of control (PF), ethanol fed (EF), probiotics fed(PF+LP), and ethanol and probiotics fed(EF+LP) microglia of mice brain cells while **(a')-(d')** are their corresponding IPR images respectively. The IPR images are distinct than the confocal images and shows that structural disorder increases in the microglia of mice brain cells in the presence of ethanol.

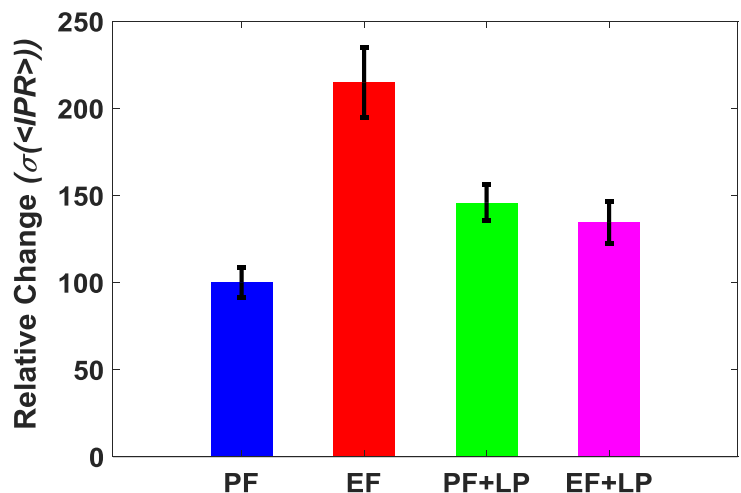

**Fig. 4:** Bar graph represents the relative study of molecular specific light localization property or $\sigma(<IPR>)$ of microglia, brain cells of control (PF) mice with respect to ethanol fed (EF), probiotic fed (PF+LP), and ethanol and probiotics fed (EF+LP) mice. The IPR analysis of microglia shows that the *std* of disorder strength, $\sigma(<IPR>)$ of EF mice increase by 120% in reference to PF mice. The $\sigma(<IPR>)$ of microglia of mice fed with only probiotics is relatively higher than the control mice. This increase in aggressiveness indicates that there is some effect of probiotics in microglia of the brain cells. However, the $\sigma(<IPR>)$ of microglia cells of mice fed with ethanol and probiotics at the same time decreases by 100% relative to ethanol fed mice. This implies probiotics in the presence of ethanol lead to more stimulated and soothed brain cells, reducing the $\sigma(<IPR>)$ of microglia of the brain cells.

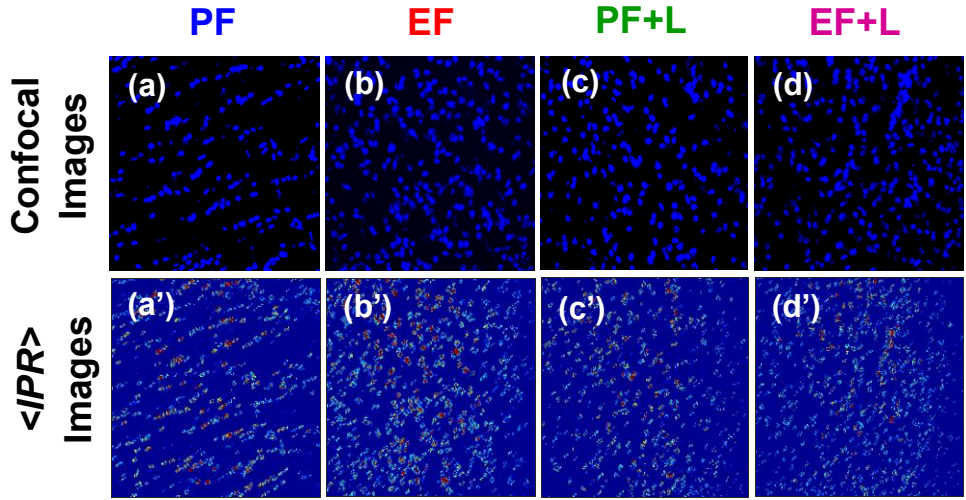

**Fig: 5. Confocal and IPR images of nuclei chromatin (PF, EF, PF+LP, EF+LP):** (a)-(d) are the representative confocal images of control (PF), ethanol fed(EF), probiotics fed(PF+LP), and ethanol and probiotics fed(EF+LP) of chromatin nuclei in the mice brain cells while (a')-(d') are their corresponding IPR images respectively. The IPR images are distinct than the confocal images and show that the structural disorder increases in the nuclei of mice brain cells in the presence of ethanol.

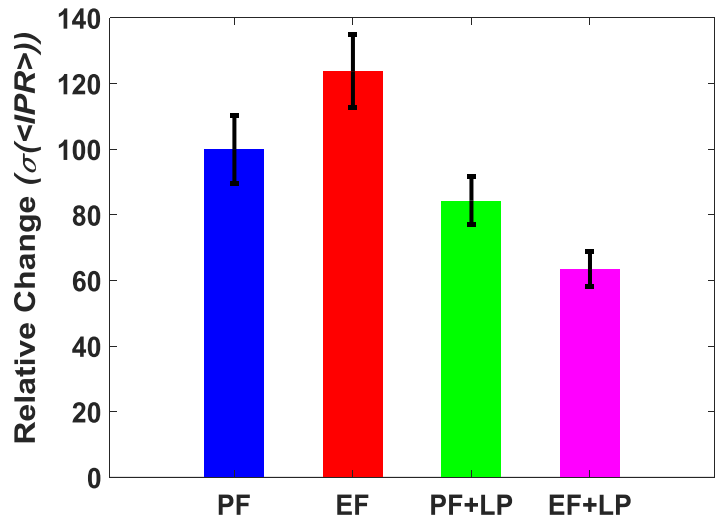

**Fig. 6.** Bar graphs represents the relative values of DNA molecular specific (or chromatin) light localization property or $\sigma(<IPR>)$ of nuclei in the brain cells of control (PF) mice with respect to ethanol fed (EF), probiotic fed (PF+LP), and ethanol and probiotic fed (EF+LP). The IPR analysis of nuclei in the brain cells of mice shows that the *std* of disorder strength, $\sigma(<IPR>)$ of EF mice increase by 25% relative to PF mice. The $\sigma(<IPR>)$ of nuclei in the brain cells of mice fed with probiotics only is relatively lower compare to the control mice. This decrease in aggressiveness indicates that there is some effect of probiotics in nuclei of the brain cells structure. However, the $\sigma(<IPR>)$ of nuclei in mice brain cells fed with ethanol and probiotics at the same time decreases by 40% relative to EF mice. This implies probiotics in the presence of ethanol are more active and help in soothing brain and reducing the $\sigma(<IPR>)$ of nuclei in the brain cells.

The representative confocal images of astrocytes from the thin section of mice brain cells fed with: (i) Control Fed (PF), (ii) Ethanol Fed (EF), (iii) Probiotics Fed (PF+LP), and (iv) Probiotics Fed simultaneous with Ethanol Fed (EF+LP) respectively represented in the above figures.

Fig.1. (a)-(d) show the confocal image of astrocytes treated GFP while Fig. 1(a')-(d') are their corresponding IPR images, respectively.

In Fig. 2, the bar graphs of relative change in the *std* of disorder strength of astrocytes in the brain cells of control (PF) mice, ethanol fed (EF), probiotic fed (LP), and ethanol and probiotic fed (EF+LP) are shown.

Fig.3. (a)-(d), similar to Fig.1, shows the confocal image of microglia treated with red dye while Fig. 3(a')-(d') are their corresponding IPR images, respectively.

Fig. 4, similar to Fig. 2, the bar graphs of relative change in the *std* of disorder strength of microglia cells of control (PF) mice, ethanol fed (EF), probiotic fed (LP), and ethanol and probiotic fed (EF+LP) are shown.

Fig.5. (a)-(d), similar to Figs. 1 and 3, shows the confocal image of nuclei treated with DAPI while Fig. 3(a')-(d') are their corresponding IPR images, respectively.

Fig. 6, similar to Figs. 2 and 4, the bar graphs of relative change in the *std* of disorder strength of nuclei in the brain cells of control (PF) mice, ethanol fed (EF), probiotic fed (LP), and ethanol and probiotic fed (EF+LP) are shown.

The intensity variation in the IPR images represents structural disorder patterns in the different components of mice brain cells. Here, the higher intensity fluctuations are represented by the red color and lower intensity with blue color in all the IPR images. As can be visualized from the IPR images, the *std* of disorder strength or $\sigma(<IPR>)$ increases in astrocytes, microglia, and chromatin in brain cells fed with ethanol relative to control indicates that alcohol has an adverse effect in the brain cells. However, when fed with only probiotics, the probiotics have some sort of effect on the brain so the disorder strength is slightly higher than the normal. This increase in the disorder strength of glial cells astrocytes, and microglia, and chromatin of the brain cells in the presence of probiotics only and go reaction to the brain cells to some extent. When the mice were fed with both probiotics and ethanol simultaneously, the $\sigma(<IPR>)$ decrease back to normal or less than the normal, suggesting probiotics drug functions well in the presence of alcohol and increases the efficiency of the brain cells. Probiotics are considered good for brain cells and help to soothe and increase the cognitive function of the brain.

As can be seen from Figs. 1 and 2, the relative study of $\sigma(<IPR>)$ shows that the structural disorder of astrocytes brain cells increases when mice were fed with ethanol, whereas it decreases significantly when mice were subsequently fed with probiotics and alcohol at the same time. Also, IPR analysis shows that the probiotics themselves have some interaction with the astrocytes and increases the structural disorder of astrocytes in the brain cells. The $\sigma(<IPR>)$ of astrocytes brain cells of mice fed with only probiotics is relatively higher compared to the PF. Although probiotics are believed to boost mood and increase cognitive function of the brain, the result shows the brain can be reactive to some specific components of brain cells like astrocytes. In particular, the bar graphs show that the $\sigma(<IPR>)$ of astrocytes in the brain cells of EF mice increase by 280% relative to PF mice. This implies that alcohol has an adverse effect in the brain cells, especially in astrocytes which perform a variety of tasks such as axon guidance, maintenance of redox potential, regulation in neurotransmitter and ion concentrations, synaptic support, blood flow control, removal of toxins and debris from the cerebrospinal fluid, etc. [28]. Astrocytes, being the most numerous cells within the central nervous system (CNS) of the brain, chronic alcoholism exacerbate neuronal dysfunction and advances mechanisms in potentiating or nullifying the pathway of neuropathologic injury [29]. On the other hand, the $\sigma(IPR)$ of astrocytes cells decrease by 100% relative to EF mice when the mice were fed with probiotic and ethanol at the same time. This decrease in the $\sigma(<IPR>)$ of astrocytes cells of EF mice supports that probiotics in the presence of ethanol are good for brain cells and help in maintaining proper brain functions.

In Figs. 3 and 4, molecular specific light localization, the IPR analysis was performed in microglia, neuronal support cells present in the CNS which primarily function as the immune system of the brain. The relative study of the $\sigma(<IPR>)$ of microglial cells of PF mice with EF, PF+LP, and EF+LP are presented. The IPR analysis performed at the same sample length of all cases shows that the $\sigma(<IPR>)$ of EF mice increases by 120% relative to PF mice. This increase in the *std* of disorder strength of microglia brain cells of EF mice suggests that alcohol has a negative effect on the immune system of brain cells. Microglia are considered as the macrophages of the CNS which clean up the cellular debris and participate in neuroinflammation to various intrinsic and extrinsic stimuli. In addition to well established phagocytic function, and innate immune function microglia involve in the development of CNS immunopathology [30]. It is known that alcohol enhances immunomodulatory molecules such as corticosterone and endotoxin which degrades the neuroimmune cells of the brain and selectively modulates the intracellular signal transductions of microglia [31]. The IPR analysis shows that $\sigma(<IPR>)$ of microglial cells of mice fed with only probiotics is relatively higher than in PF mice. This might be due to the interaction of probiotics with microglia brain cells in an adverse way resulting in increasing the degree of disorder strength. Probiotics used to soothe and increase the cognitive function of the brain may be sometimes reactive and harmful to the neuroimmune system of the brain cells; however, the $\sigma(<IPR>)$ of microglia brain cells of mice fed with alcohol and probiotics simultaneously decreases by 100% relative to EF mice. That means probiotics have increased efficacy in the presence of alcohol which helps in reducing the $\sigma(<IPR>)$ or disorder strength of microglia brain cells. Therefore, probiotics with alcohol are good for brain cells and enhance the development of CNS immunopathology.

In Figs. 5 and 6, the color maps and bar graphs of relative change in the disorder strength of chromatin in brain cells of PF mice with respect to EF, PF+LP, and EF+LP are shown. The IPR analysis of DAPI stained confocal images of mice brain cells at sample length *l nm* shows that the $\sigma(<IPR>)$ of EF mice increase by 25% in reference to PF mice. This increase in the $\sigma(<IPR>)$ is due to the adverse effect of alcohol in the chromatin of brain cells. The increase in mass density fluctuations of the chromatin due to alcohol is responsible for an increase in the *std* of disorder strength. Persistent alterations to the chromatin structure are factors for epigenetics inheritance and can have a long-lasting influence on the activity and connectivity functions of the brain [32]. The $\sigma(<IPR>)$ of nuclei in the brain cells of mice fed with only probiotics is relatively lower compared to the PF mice indicate that probiotics are good and soothe the chromatin of brain cells to some extent with the increase in chromatin's efficiency. Further, the $\sigma(<IPR>)$ of nuclei in brain cells of mice fed with alcohol and probiotics at the same time decreases by 40% relative to ethanol fed mice. This decrease in the disorder strength of the chromatin of mice fed with alcohol and probiotics at the same time is due to the fact that probiotics in the presence of alcohol get more stimulated allowing them to function better which are good for the nucleus of the brain cells.

The nanoscale quantification of alcohol effect on brain cells using the IPR analyses shows that alcohol has an adverse effect on astrocytes, microglia, and chromatin. Probiotics alone are also responsible for an increase in the *std* of disorder strength in astrocytes and microglia cells while a decrease in the *std* of disorder strength in the nuclei of the brain cells. Probiotics alone affect the astrocytes, microglia, and nucleus of the brain cells differently. However, the results show that probiotics in the presence of ethanol get stimulated and function better in the brain cells, resulting in soothing the cell structure and increasing to the cognitive function of the brain.

### 4. Conclusion

The molecular specific light localization technique, Confocal-IPR is used to study the effect of probiotics in chronic alcoholism in brain cells and chromatin using a confocal image and a mouse model. The nanoscale mass density fluctuations quantified by *std* of the average IPR show an increase in disorder strength fluctuations or $\sigma(<IPR>)$ from control fed to ethanol fed mice in astrocytes, microglia, and chromatin of the brain cells. This increase in structural disorder confirms the adverse effect of alcohol in astrocytes, microglia, and chromatin of the brain cells. On the other hand, an increase in the disorder strength of brain cells when control mice were fed with probiotics to improve the efficacy of brain might be due to brain cells reactions which are excited in the presence of probiotic drugs in some specific cells such as astrocytes and microglia. In detail, the study might be required to elucidate plausible reasons for the increase in $\sigma(<IPR>)$ and have adverse effects in astrocytes and microglia although probiotic drugs are used to improve cognitive function of the brain. Interestingly, decrease in the $\sigma(<IPR>)$ in astrocytes, microglia, and chromatin of the brain cells fed with ethanol and probiotics simultaneously relative to normal and ethanol fed indicate that probiotics in the presence of alcohol get highly stimulated and helps in soothing the brain cells physical structure and increasing the multifunctionality of brain. That means the astrocytes, microglia, and chromatin which includes a major portion of the CNS of the brain and perform vital brain functions such as blood flow control, removal of toxins and debris from the cerebrospinal fluid, immune system, genetic inheritance, neuroinflammation to various intrinsic and extrinsic stimuli, etc. get improved. The almost reversible effect of probiotics given with the alcohol in the brain cells could be a useful way to mitigate or improve abnormalities in the brain cells due to alcohol, stress or any other harmful sedative drugs, a major concern of modern life. As an illustration of the potential application of the mesoscopic physics-based IPR technique, we have successfully measured the structural alterations at the nanoscale level in the mice brain cells and chromatin due to probiotics in chronic alcoholism.

**Acknowledgments**

PPradhan acknowledges NIH and MSU for financial supports.